

\input eplain

\newcount\fignumber
\def\figdef#1{\global\advance\fignumber by 1 \definexref{#1}{\number\fignumber}{figure}\ref{#1}}
\def\figdefn#1{\global\advance\fignumber by 1 \definexref{#1}{\number\fignumber}{figure}}
\let\figref=\ref
\let\figrefn=\refn
\let\figrefs=\refs

\newcount\tabnumber
\def\tabdef#1{\global\advance\tabnumber by 1 \definexref{#1}{\number\tabnumber}{table}\ref{#1}}
\def\tabdefn#1{\global\advance\tabnumber by 1 \definexref{#1}{\number\tabnumber}{table}}

%
\ifx\pdfoutput\undefined
\input epsf

\def\figscale#1#2{\epsfxsize=#2\epsfbox{#1.eps}}
%
\else

\def\figscale#1#2{\pdfximage width#2 {#1.pdf}\pdfrefximage\pdflastximage}
\fi


\newcount\scount \scount=0



\makeatletter
\def\section#1\par{
  \vskip\z@ plus.3\vsize\penalty-250
  \vskip\z@ plus-.3\vsize\bigskip\vskip\parskip
  \global\advance\scount by1
  \writenumberedtocentry{section}#1{}
  \definexref#1{\the\scount}{section}
  \message{#1}
  \noindent\the\scount.\quad{\bf #1}\nobreak\smallskip\noindent}
\makeatother

\centerline{\bf{Fractal Scaling of Population Counts Over Time Spans}}
\centerline{Aubrey G. Jaffer and Martin S. Jaffer}
\centerline{Digilant}
\centerline{2 Oliver Street, Suite 901}
\centerline{Boston MA 02109 US}
\centerline{agj@alum.mit.edu}

\beginsection{Abstract}

{\narrower

        Attributes which are infrequently expressed in a population
        can require weeks or months of counting to reach statistical
        significance.  But replacement in a stable population
        increases long-term counts to a degree determined by the
        probability distribution of lifetimes.

        If the lifetimes are in a Pareto distribution with shape
        factor $1-r$ between 0 and 1, then the expected counts for a
        stable population are proportional to time raised to the $r$
        power.  Thus $r$ is the fractal dimension of counts versus
        time for this population.

        Furthermore, the counts from a series of consecutive
        measurement intervals can be combined using the $L^p$-norm
        where $p=1/r$ to approximate the population count over the
        combined time span.

        Data from digital advertising support these assertions and
        find that fractal scaling is useful for early estimates of
        reach, and that the largest reachable fraction of an audience
        over a long time span is about $1-r$.

\par}
\bigskip
{\noindent {\bf Keywords}: population statistics; fractal dimension; digital advertising}

\beginsection{Table of Contents}

\readtocfile

\section{Introduction}

In online (digital) advertising, ``third-party'' data vendors provide
streams of anonymized unique user identifiers (UUID) along with their
alleged attributes for use in deciding which users to buy
advertisements for.

The primary assumption here is that one UUID (in a population) being
associated with a feature is persistent and independent of other
UUIDs being associated with the feature.  This makes it a stationary
Bernoulli process (coin-toss) of $N$ trials, with expected value
$N\cdot P$, where $P$ is the probability that a UUID has a particular
attribute.

It is natural to ask what is the size of a pool of users, those with a
particular attribute or combination of attributes, and also the total
population.  But the counts of such pools depend on the time span over
which the unique identifiers are counted.

Popular web browsers offer ``incognito'' modes whose UUIDs (in the
form of cookies) are forgotten at the end of the session.  Despite the
efforts of third-party vendors to filter out ephemeral UUIDs,
short-lived UUIDs comprise the majority of the UUIDs seen
week-to-week.

Digital advertisers want to evaluate the effectiveness of targeting
any of the hundreds or thousands of third-party attributes in driving
sales.  With typical success rates of only 0.1\%, hundreds of
advertisements must be bought per sale.  In order to achieve
statistically significant measurements, the UUID counts for attributes
and combinations of attributes must span weeks.

Counting unique cookies over a 6 month span can be expensive in
computing and storage costs.  This investigation began as a study of
the relationship between weekly counts and counts over multiple weeks.

\section{The $L^p$-norm}

Consider the population counts for two consecutive weeks and a
two-week count for the same time period.  Every individual counted in
the weekly counts must also appear in the two-week count and every
individual counted in the two-week count must appear in at least one
of the weekly counts.  So these three counts must obey the triangle
inequality.

The triangle inequality suggests that the weekly counts might be
treated as dimensions, and the two-week count as the result of a norm
applied to their vector sum.  If the population is very long-lived,
then few individuals get replaced, and the population count will be
nearly constant with time.  If the individuals in a stable population
are short-lived, then the population count will grow nearly linearly
with the duration of the count.  Experimentation with the graphs
quickly converged to the distinct $p$ exponents in \ref{Digital
Advertising Data}, which worked so well that it prompted this
exploration of the mathematics.

The $L^p$-norm is:

$$\left\|C_1,\dots,C_n\right\|_p=\left(|C_1|^p+\dots+|C_n|^p\right)^{1/p}
=\left(\sum_{j=1}^n|C_j|^p\right)^{1/p}$$ 

All population counts $C_j$ are non-negative, so the absolute values
are superfluous to this application.

With $p=1$ the weekly counts add linearly.  As $p$ approaches
$\infty$, the $L^\infty$-norm returns the maximum of its inputs.
These limits satisfy the earlier reasoning.

The $L^p$-norm is idempotent; input $C$ values can be combined without
changing the resulting value:


$$\eqalign{\left\|\left\|C_1,C_2\right\|_p,C_3,\dots,C_n\right\|_p
 &=\left(\left(\left(|C_1|^p+|C_2|^p\right)^{1/p}\right)^p+|C_3|^p+\cdots+|C_n|^p\right)^{1/p}\cr
 &=\left(\left(|C_1|^p+|C_2|^p\right)+|C_3|^p+\cdots+|C_n|^p\right)^{1/p}\cr
 &=\left\|C_1,C_2,C_3,\dots,C_n\right\|_p
}$$                             

With the assumption that $p$ remains constant over time, the graphs
in \ref{Digital Advertising Data} show that we can estimate the population over a span
of $n$ weeks from population counts of each of the constituent weeks
using the $L^p$-norm.

The $L^p$-norm definition implies a scaling law.  If all the $C_j$
have the same value $C$, then:

$$\left\|C_1,\dots,C_n\right\|_p=\left(\sum_1^n\left|C\right|^p\right)^{1/p}
         =\left(n\cdot|C|^p\right)^{1/p}=C\cdot n^{1/p}
\eqdef{power-derivation}$$

The norm for the analogous $L^p$ space is:

$$\left\|C\right\|_p\equiv\left(\int_{0}^t|C(t)|^p\,dt\right)^{1/p}
 \eqdef{continuous}$$

When $C$ is constant, norm \eqref{continuous} obeys the same scaling
law as the $L^p$-norm \eqref{power-derivation}.

\section{Fractal Dimension}

On viewing scaling law \eqref{power-derivation}, the authors realized
that $r=1/p$ is a fractal dimension (as described by
Mandelbrot\cite{Mandelbrot:1977:FFC}).  Similarly to the length of a
coastline growing as the measurement resolution is increased, the
count of a population increases as the time span of counting
increases.

This implied scaling law \eqref{power-derivation} is plotted along
with the true and $L^p$-estimated counts in the graphs
in \ref{Digital Advertising Data}.  It is in rough agreement with the multiple-week
counts and $L^p$-estimates, even though there is some variation in the
weekly counts.

\section{Digital Advertising Data}

The figures show the UUIDs per week, the $L^p$-norm of $n$ weekly
counts, the UUIDs counted over a span of $n$ weeks, and the scaling
law with its coefficient being the geometric mean of the weekly
counts.

 \figref{ogrowth} shows a 7-week span starting in March 2016 of all
 UUIDs seen by a large third-party vendor.  The variation in the
 number of UUIDs per week is tracked well by the $L^{1.65}$-norm; less
 so by the scaling law with its assumption of identical weekly counts.

\vbox{\settabs 1\columns
\+\hfill\figscale{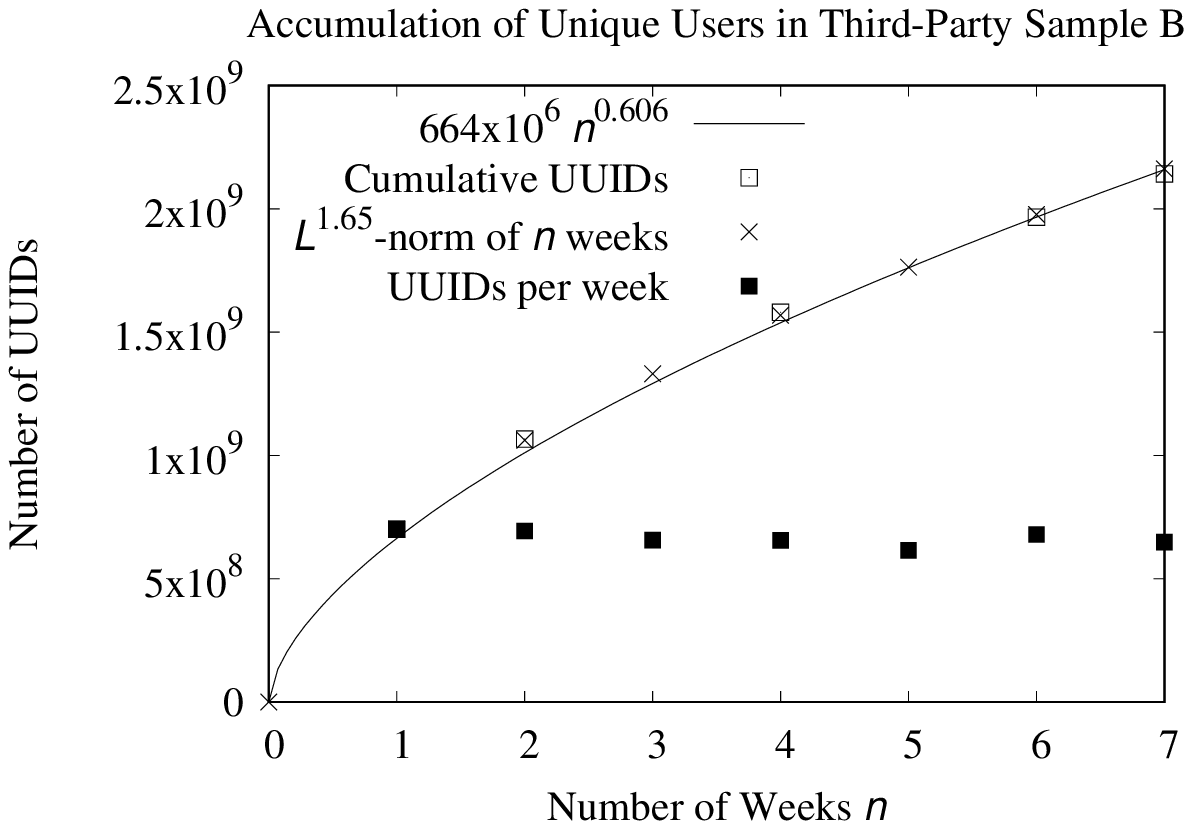}{300pt}\hfill&\cr
\+\hfill\figdef{ogrowth}\hfill&\cr
}

 \figref{growth} shows a 14-week span starting 2017-11-20 of all UUIDs
 seen by another third-party vendor.  This vendor provides some
 attributes which depend on how many times a UUID clicks, which
 violates the assumption of a stationary Bernoulli process.

\vbox{\settabs 1\columns
\+\hfill\figscale{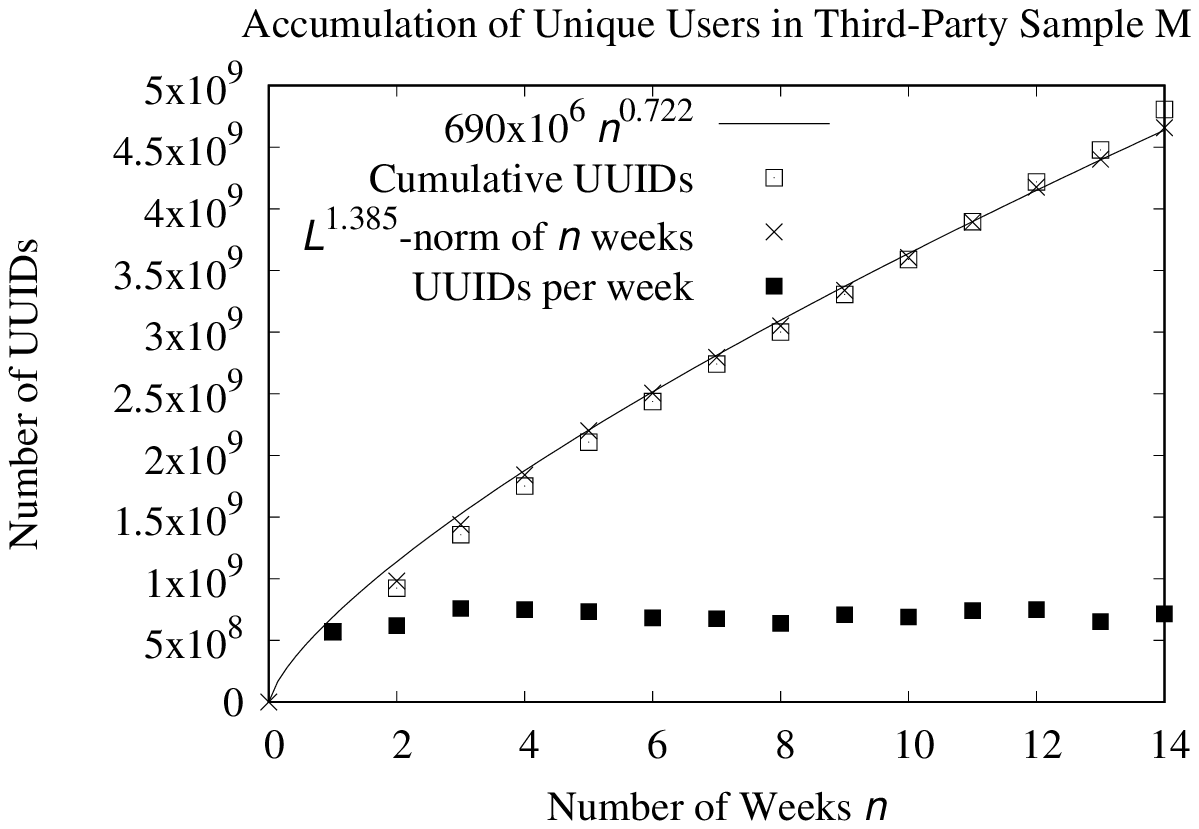}{300pt}\hfill&\cr
\+\hfill\figdef{growth}\hfill&\cr
}

\vfill\eject

 \figref{pgrowth} shows a 14-week span starting 2017-11-20 of all
 UUIDs seen by Digilant advertisers' pixels\numberedfootnote{In
 digital advertising a pixel is a tiny image used to learn the UUIDs
 of visitors to a web-page containing that pixel.}.  Being unfiltered,
 this data-set has a fractal dimension larger than 0.92.

\vbox{\settabs 1\columns
\+\hfill\figscale{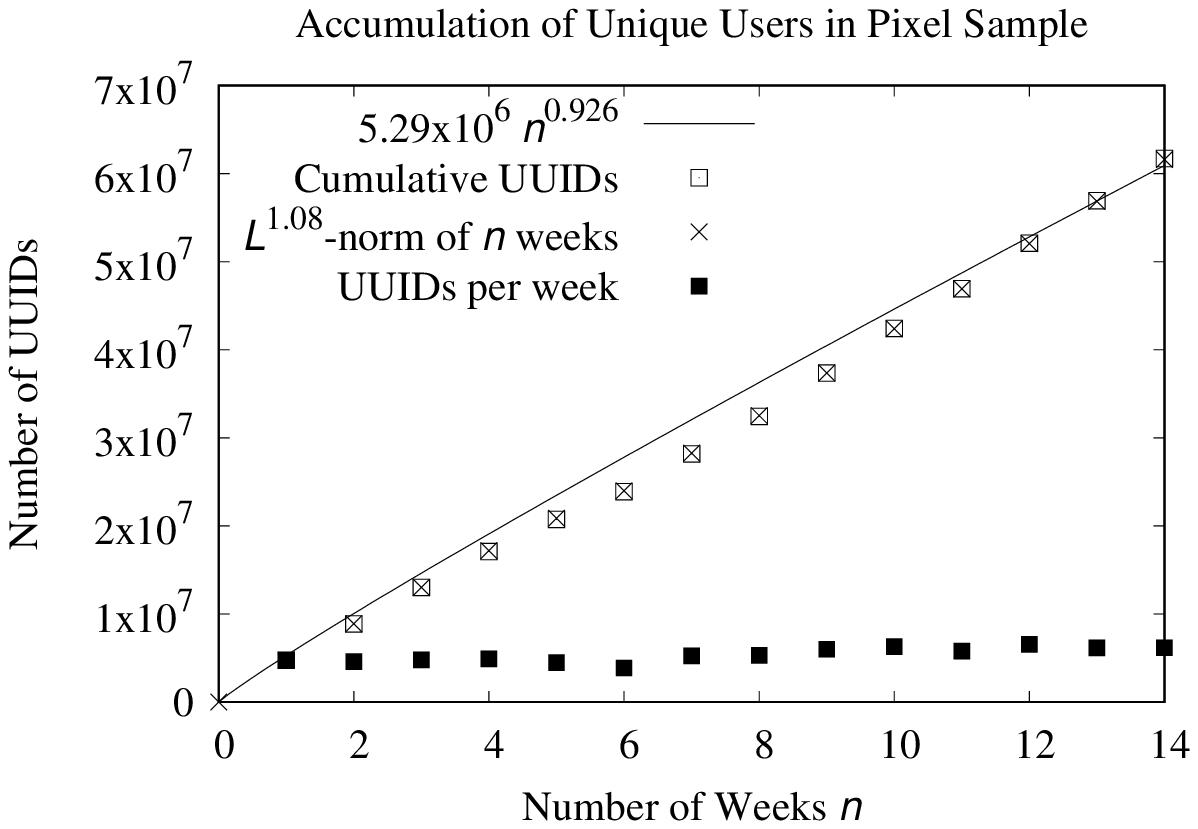}{300pt}\hfill&\cr
\+\hfill\figdef{pgrowth}\hfill&\cr
}

 \figref{pixdy} shows a 31-day span starting 2017-11-20 of the same
 pixels hits.  The cumulative counts are in close agreement with
 $L^{1.085}$-norm counts.

\vbox{\settabs 1\columns
\+\hfill\figscale{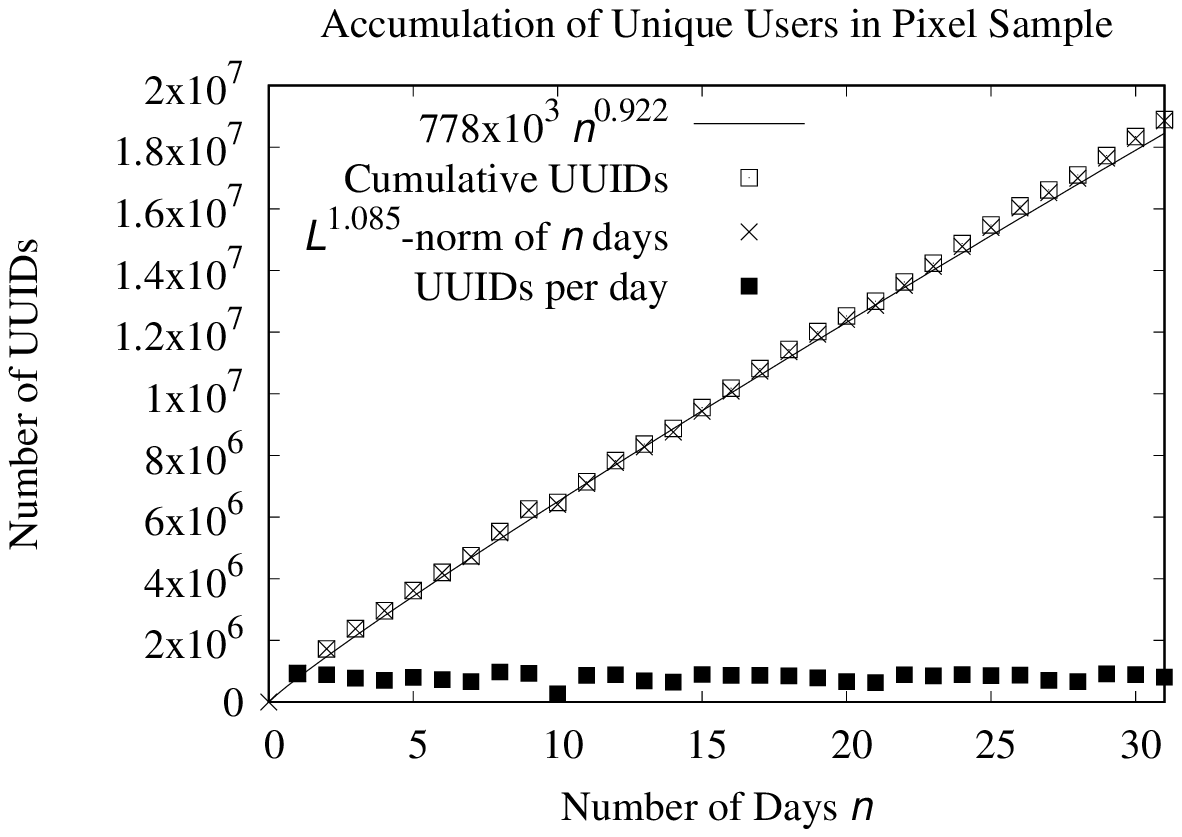}{300pt}\hfill&\cr
\+\hfill\figdef{pixdy}\hfill&\cr
}

\section{Asymptotics}

Assume a stable population of size $n$ and time $t\ll n$.

In order to uniquely count the population over time $t$, the storage
required is $O(t^r\,n\,\log n)$; and the running time is $O(t\,n\,\log
n)$.  If cumulative counts are to be computed every time period, then
the storage is $O(t\,n)$ and running time is $O(t^2\,n\,\log n)$

If instead, counts are made every time unit (to be combined with the
$L^p$-norm), then the running time is $O(t\,n\,\log n)$, the
short-term storage is $O(n)$ and the long-term storage is $O(t)$

\section{Pareto Distribution}

Is there a probability distribution for lifetimes which produces
$L^p$-norm and fractal scaling of population counts?  Going through
McLaughlin's {\it ``A compendium of common probability
distributions''}\cite{mclaughlin:dist}, it was found that the Pareto
probability distribution, which is used for modeling income and
longevity distributions, has the fractal scaling properties.

Let $X$ be a Pareto random variable for UUID lifetime with positive
scale factor $A\le X$ and positive shape factor $B<1$.

$$p_X(x)=P(X=x)={BA^B\over x^{B+1}}\qquad P(X<x)
        =1-\left({A\over x}\right)^B$$

With a stable population, each individual is replaced when its
lifetime $X$ expires.  The count of replacements over time $t$ is:

$$R(t)={N\,t\over X}$$

For $t\ge1$ let $A=1/t$.  Ignoring cohorts with average lifetimes
shorter than the unit time interval, the expected replacement count
is:

$$E[R(t)]=\int_{1}^{\infty} {N\,t\cdot p_X(x)\over x} dx
  =Nt \int_{1}^{\infty} {B\,A^{B}\over x^{B+2}} dx
  =N{B\over B+1}\,t^{1-B} \eqdef{E[R(t)]} $$

With $r=1-B$, the expected count \eqref{E[R(t)]} comes into the same
fractal scaling form as equation \eqref{power-derivation}:

$$E[R(t)]=N{B\over B+1}\,t^{1-B}=N{1-r\over 2-r}\,t^r=E[R(1)]\,t^r$$

The expected population count $E[C(t)]=N\,t^r$ is proportional to the
expected replacement count:

$$E[C(t)]={2-r\over 1-r}\,E[R(t)]\qquad E[C(t)]=E[C(1)]\,t^r\eqdef{t^r}$$

Given the fractal scaling of expected count \eqref{t^r}, what can be
inferred about splitting $E[C(t)]$ into $t$ equal size counts
$C=C(1)$?  From equation \eqref{t^r} the unknown function
$f(C,\dots,C)=C\cdot t^r$.  Raising both sides to the $1/r$ power:

$$f(C,\dots,C)^{1/r}=t\cdot C^{1/r}=C^{1/r}+\cdots+C^{1/r}$$

  Raising both sides to the $r$ power:

$$f(C,\dots,C)=\left(C^{1/r}+\cdots+C^{1/r}\right)^r\eqdef{raise}$$

The right side of equation \eqref{raise} is the formula for the
$L^p$-norm $\left\|C_1,\dots,C_t\right\|_{1/r}$ for non-negative
$C_j$.  Thus the Pareto lifetime distribution implies a scaling law,
which in turn implies the $L^p$-norm with $p=1/r$ for successive
non-overlapping counts.

\section{Changing Population Size}

So far we have assumed that population sizes did not experience much
increase or decrease during the measurement interval.  While the
triangle inequality holds when the constituent counts are very
different in magnitude, a population cannot drop to zero in one time
period without invalidating the longer lifetimes in the probability
distribution of the previous period.

However, the $L^p$-norm estimates of sample B (\figref{ogrowth}) and
pixel (\figrefs{pgrowth} and \figrefn{pixdy}) populations track the
cumulative counts well through variations in the weekly and daily
counts (over 3.5:1 in the daily pixel case).





\section{Measuring the Fractal Dimension}

Given positive daily population counts $C_m,\dots,C_n$ and
corresponding (monotonically increasing) cumulative counts
$Q_m,\dots,Q_n$, for $0<m\le j\le n$ we would like to find the optimal
$p$ to minimize the difference between
$\left\|C_1,\dots,C_j\right\|_p$ and $Q_j$.  Equivalently, we wish to
minimize the difference between $Q_j^p-Q_{j-1}^p$ and $C_j^p$ (where
$Q_0=0$).


Suppose we have an initial value for $p$ which does not extinguish the
difference between $Q_j^p-Q_{j-1}^p$ and $C_j^p$.  Let $\delta$ be the
change in exponent $p$ which makes them equal:

$$Q_j^{p+\delta}-Q_{j-1}^{p+\delta}=C_j^{p+\delta}$$
$$Q_j^{\delta}Q_j^{p}-Q_{j-1}^{\delta}Q_{j-1}^{p}=C_j^{\delta}C_j^{p}$$

If $\delta$ is near zero and $Q_j^{\delta}$ and $Q_{j-1}^{\delta}$ are
close in value, then they can be approximated by their average
$Q_{j'}={(Q_{j}+Q_{j-1})/2}$.

$$Q_{j'}^{\delta}\left(Q_j^{p}-Q_{j-1}^{p}\right)\approx C_j^{\delta}C_j^{p}$$
$${Q_j^{p}-Q_{j-1}^{p}\over C_j^{p}}\approx {C_j^{\delta}\over Q_{j'}^{\delta}}$$
$$\log\left({Q_j^{p}-Q_{j-1}^{p}\over C_j^{p}}\right)\approx \delta_j\log\left({C_j\over Q_{j'}}\right)$$
$$\delta_j\approx
 {\log\left(\left[Q_j^{p}-Q_{j-1}^{p}\right]/C_j^{p}\right)\over
 \log\left({2\,C_j/\left[Q_j+Q_{j-1}\right]}\right)}$$

By averaging $\delta$ over $j$, $p$ can be improved for the dataset as
a whole:

$$\delta=
 {1\over n-m+1}
 \sum_{j=m}^n
 {\log\left(\left[Q_j^{p}-Q_{j-1}^{p}\right]/C_j^{p}\right)\over
 \log\left({2\,C_j/\left[Q_j+Q_{j-1}\right]}\right)}
 \qquad p\gets\delta+p$$

Overshoot from $p\gets\delta+p$ leads to slow oscillatory convergence.
$p\gets0.632\,\delta+p$ converges about one decimal digit per
iteration.  Once $p$ has settled, its standard-deviation can be
calculated:

$$\sigma=\sqrt{
 {1\over n-m+1}
 \sum_{j=m}^n
 \left[{\log\left(\left[Q_j^{p}-Q_{j-1}^{p}\right]/C_j^{p}\right)\over
 \log\left({2\,C_j/\left[Q_j+Q_{j-1}\right]}\right)}\right]^2}$$

In practice, the contribution from step $j$ should only be included
when $0<1.6\,C_j<\left(Q_j+Q_{j-1}\right)/2$; the count $n-m+1$ is
reduced by the number of excluded steps.

\section{Reach and Saturation}

Reach is the total number of individuals who received or viewed an
advertisement during the campaign; reach goals are often part of
advertising contracts.  Fractal scaling provides good early estimates
of large reaches, as can be seen from the graphs in \ref{Digital
Advertising Data}.

The $p$ for daily reach from more than fifty Digilant managed
campaigns in the month of June 2018 were between 1.002 and 1.24 with a
mean of 1.074; the weekly $p$ for the same time period were between
1.007 and 1.20 with a mean of 1.077.\numberedfootnote{The only
campaign which was purely retargeting (repeatedly showing
advertisements to the same users) during June had a $p$ of 1.79 and
was not included in the averages.}  That these averages (1.074 and
1.077) are so close to the $p$ for pixel hits (1.085 and 1.08), is
evidence that the fractal dimension is an intrinsic property of the
user population.

Saturation is the ratio of the reach to the number of UUIDs with the
targeted attributes.  For UUID populations with long lifetimes, this
ratio can approach 1.  The ratio is small for populations with short
lifetimes because the UUIDs tend not to be online long enough to see
many advertisements.

Looking at 6 months of Digilant advertising campaigns which targeted
attributes from samples B and M, the largest saturation achieved (for
sample B and for sample M) was roughly $1-r$.

\section{Conclusion}

For populations having a Pareto distribution of lifetimes with shape
factor $0<B<1$, counts made over successive time intervals can be
combined using the $L^p$-norm to closely approximate the count which
would result from counting over the combined time span.  The norm's
exponent $p=1/r$ where $r=1-B$ is the fractal dimension of the
population counts over time.  Fractal scaling allows counts collected
over very different time spans to be effectively compared.

Digital advertising UUIDs are an example of such a population.
Collecting daily or weekly counts, then aggregating using the
$L^p$-norm, allows longer term studies with better confidence to be
conducted without straining resources.

Fractal scaling laws imply aggregation using the $L^p$-norm.  Regions
where the norm doesn't scale with the expected exponent might be used
to locate anomalies in large temporal or spacial data-sets.

\section{References}

\bibliographystyle{unsrt}
\bibliography{fracount}

\vfill\eject
\bye